\documentclass[10pt, journal]{IEEEtran}
\makeatletter
\usepackage{amssymb}
\usepackage{amsmath}
\usepackage{mathtools}
\usepackage{ellipsis}
\usepackage{textcomp}
\usepackage{cite}
\usepackage{graphicx}
\usepackage{subfigure}
\usepackage{epstopdf}
\usepackage{epsfig}
\usepackage{caption}
\usepackage{algorithm}
\usepackage{algpseudocode}
\usepackage{algorithmicx}
\usepackage{psfrag}
\usepackage{amsthm}

\theoremstyle{definition}

\theoremstyle{remark}

\theoremstyle{plain}

\usepackage{grffile}
\usepackage{setspace}
\usepackage[english]{babel}
\usepackage{caption}
\usepackage{subfigure}

\usepackage{color}

\begin{document}

\title{Quality-Driven Energy-Efficient Big Data Aggregation in WBANs
}
\author{Amit~Samanta,~\IEEEmembership{Student~Member,~IEEE}~and~Tri~Gia~Nguyen,~\IEEEmembership{Senior~Member,~IEEE}
\thanks{Amit Samanta is with the School of Computing, University of Utah, USA. (amit.samanta049@gmail.com)}
\thanks{Tri Gia Nguyen are with FPT University, Da Nang 50509, Vietnam. (tri@ieee.org)}

}

\maketitle

\begin{abstract}
In the Internet-of-Things (IoT) era, the development of Wireless Body Area Networks (WBANs) and their applications in big data infrastructure has gotten a lot of attention from the medical research community. Since sensor nodes are low-powered devices that require heterogeneous Quality-of-Service (QoS), managing large amounts of medical data is critical in WBANs. Therefore, effectively aggregating a large volume of medical data is important. 
In this context, we propose a quality-driven and energy-efficient big data aggregation approach for cloud-assisted WBANs. For both intra-BAN (Phase I) and inter-BAN (Phase II) communications, the aggregation approach is cost-effective. Extensive simulation results show that the proposed approach -- {\tt DEBA} greatly improves network efficiency in terms of aggregation delay and cost as compared to existing schemes.
\end{abstract}

\section{Introduction}
WBANs provide patients with effective real-time medical services due to significant advancements in sensor technologies. Several big data-enabled body sensors are mounted on or in the body of sensor-enabled patients in a WBAN to detect their physiological signals \cite{Zuhra2020, Ling2020}. Following that, the sensed medical data is compiled at the Local Managing Unit (LMU), which then sends the compiled data to health cloud-service providers (HCSPs) via base stations (BSs) and the Internet \cite{samanta2018dynamic}. Big data-enabled body sensors, on the other hand, generate large amounts of medical data in the terabytes or hexabytes range in a critical emergency situation \cite{ullah2019future}. Therefore, the data produced by big data-enabled body sensors follows the 3Vs (volume, variety, and velocity) of big data. The volume defines the amount of data triggered by IoT-enabled body sensors, the velocity defines the data flow rate from IoT-enabled body sensors, and the variety defines the types of data triggered by IoT-enabled body sensors (i.e., semi-structured, structured, and unstructured) \cite{wang2016big, quwaider2014efficient, du2015framework, Jamthe2015}.

Since the big data-enabled body sensor generates a large amount of data, it's crucial to effectively aggregate it. Data flow between IoT-enabled body sensors and LMUs (Phase I) and data flow between LMUs and BSs (Phase II) are the two components of the data communication phase in the WBAN architecture. As a consequence, we need to divide the large amount of produced data into two phases (i.e, Phase I and II). However, such a big data aggregation mechanism imposes several constraints for WBANs: 1) The data should be aggregated rapidly, as vital and sensitive medical data must be offloaded as soon as possible; 2) the data aggregation mechanism must be energy-efficient, as sensor nodes and LMU have limited battery capacity; and, finally, 3) the consistency of the aggregated data must be preserved to produce the desired results. Unlike the exiting works \cite{wang2016big, quwaider2014efficient, du2015framework, SamantaTMC2017, misra2018traffic}, we develop DEBA\footnote{Quality-\textbf{D}riven \textbf{E}nergy-Efficient \textbf{B}ig Data \textbf{A}ggregation for WBANs.}, a quality-driven, energy-efficient big data aggregation mechanism for WBANs to address the upper mentioned issues.

\section{System Model} 
\begin{figure}[!ht]
\centering
\includegraphics[width=8cm]{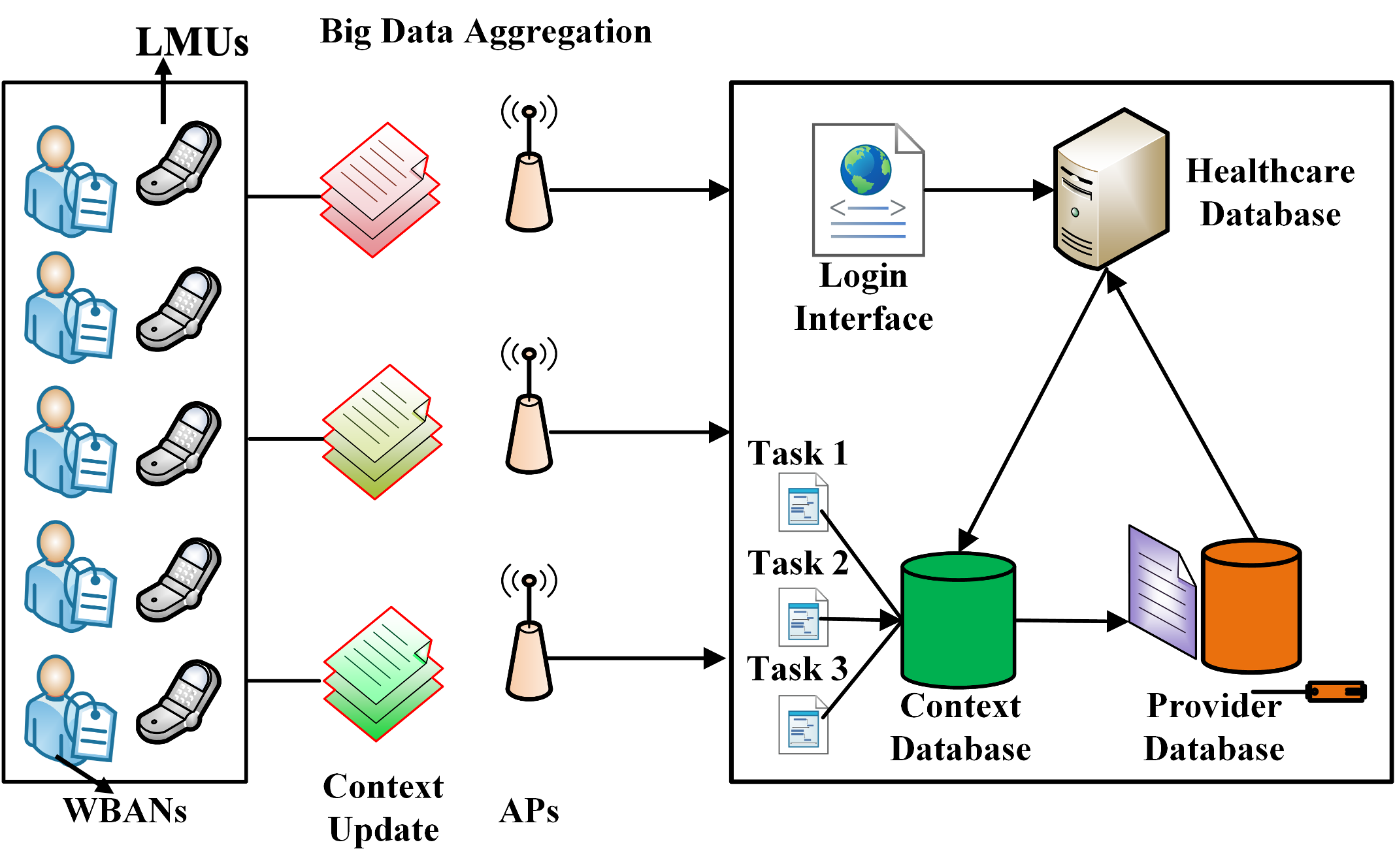}
\caption{Cloud-assisted WBANs with big data applications}
\label{fig:Arch}
\end{figure}

We consider a big data-enabled WBAN architecture in Figure \ref{fig:Arch}, where we assume $N$ WBANs \cite{Guo2020} are presented in a healthcare premise, $\mathcal{B} = \{\mathbb{B}_1, \mathbb{B}_2, \cdots, \mathbb{B}_N \}$. Each WBAN is comprised of $n$ IoT-enabled body sensors, $\mathbb{B} = \{ b_1, b_2, \cdots, b_n \}$ as shown in Figure \ref{fig:Arch}. After sensing the physiological data, IoT-enabled body sensors transmit the accumulated health specifications to LMUs and LMUs transmit different health related variables to BSs. However, the aggregation of huge data at the LMUs from heterogeneous big data-enabled body sensors is very challenging due to its limited processing power and buffer size. Therefore, it is necessary to propose an efficient and cost-effective big data aggregation process for cloud-assisted WBANs, while preserving the \textit{fair}-QoS distribution process among WBANs to increase network performance.     
 
The big data aggregation cost for big data-enabled WBANs is expressed as:
\begin{equation}
\mathcal{Y}_r = \sum_{i=1}^n \sum_{k=1}^k \gamma p_k f(w_{agg}^i)
\end{equation}
where $\gamma$ represents the health data aggregation factor, $p_k$ represents the data aggregation cost for aggregation of $k$ chunks of data, and $f(w_{agg}^i)$ denotes the big data aggregation function for cloud-assisted WBANs. Here, $f(w_{agg}^i)$  is represented as the weights $w_{agg}^i$ to aggregate the big data in WBANs and weights are set in the range of $[0,1]$. The QoS-constraint for big data aggregation is mathematically expressed as:
\begin{equation}
\mathcal{U}_r(\mathcal{Y}_r) = \sum_{t=1}^{\mathcal{T}} \bigg( \frac{\mathcal{S}_{req}^t - \mathcal{S}_{pre}^t }{ \mathcal{S}_{req}^t } \bigg) \mathcal{Y}_r
\end{equation}
where $\mathcal{S}_{req}^t$ and $\mathcal{S}_{pre}^t $ denote the required QoS and present QoS of WBANs for big data aggregation process at time $t$. $\mathcal{T}$ is represented as the total time frame for big data aggregation and it is divided into small epoch $t$. Therefore, our objective is to provide energy-efficient and cost-effective big data aggregation process for quality-driven cloud-assisted WBANs.   

\section{Quality-Driven Data Aggregation}
In a critical emergency situation, the big data-enabled body sensors generate a huge amount of medical data for the early detection of medical diseases. Therefore, it is very challenging for LMUs to aggregate medical data in an efficient way. In this context, we proposed a quality-driven data aggregation approach to aggregate the huge amount of big data in LMUs.  

The data aggregation model for cloud-assisted WBAN is represented as:
\begin{equation}
\mathcal{Z}_i =  \sum_{j=1}^N \sum_{i=1}^n \mathcal{H}_{ij} \bigg( \mathcal{Z}_t + r_i (1- \mathcal{L}_{ij}) \bigg)
\end{equation} 
where $\mathcal{H}_{ij}$ is set to $1$ if there exists a link between $i$ and $j$ or set to $0$, $\mathcal{Z}_t$ denotes aggregated traffic at each sensor node for a duration of $t$, $r_i$ represents the big data transmission rate of body sensor $b_i$, and $\mathcal{L}_{ij}$ denotes the failure probability of big data generated from IoT-enabled body sensors. 

The data aggregation function for $i^{th}$ WBAN is dependent on the delay function, energy constraint, and buffer utilization function of data aggregation model, which is expressed as: 
\begin{equation}
f(w_{agg}^i) = \sum_{t=1}^{\mathcal{T}} \bigg( f(w_{delay, t}^i) + f(w_{energy, t}^i) + f(w_{buff, t}^i) \bigg) \mathcal{Z}_i^t
\label{main}
\end{equation}
The delay function determines the incurred data aggregation delay for the cloud-assisted delay, which is defined as:
\begin{equation}
f(w_{delay, t}^i) = \bigg[ \frac{\mathcal{D}_{t, prop}^i + \mathcal{D}_{t, tran}^i + \mathcal{D}_{t, agg}^i}{ \mathcal{D}_{t, tot}^i}\bigg] 
\end{equation}
where $\mathcal{D}_{t, prop}^i$, $\mathcal{D}_{t, tran}^i$, $\mathcal{D}_{t, agg}^i$, and $\mathcal{D}_{t, tot}^i$ denote the propagation delay, transmission delay, data aggregation delay, and total delay incurred at big data-enabled body sensor $b_i$, respectively. The energy constraint for data aggregation in cloud-assisted WBAN is expressed as:
\begin{equation}
f(w_{energy}^i) = \mathcal{E} \frac{E_{tran, t}^i}{E_{ini, t}^i}
\end{equation} 
where $E_{tran,t}^i$, $E_{ini,t}^i$, and $\mathcal{E}$ denote the present residual energy, initial residual energy, and scaling factor, respectively. The buffer utilization function is expressed as:
\begin{equation}
f(w_{buff}^i) = \bigg( \frac{\mathcal{Q}_{tot_i} - \mathcal{Q}_{pre_i}^t}{\mathcal{Q}_{tot_i}} \bigg)
\end{equation} 
where $\mathcal{Q}_{tot_i}$ and $\mathcal{Q}_{pre_i}^t$ denote the total buffer size of a big data-enabled body sensor $b_i$ and present buffer size of a big data-enabled body sensor $b_i$ at time $t$. The quality driven big data aggregation in cloud-assisted WBANs is defined as:
\begin{align} \notag
 \sum_{t=1}^\mathcal{T} \zeta_t = \bigg( f(w_{delay}^i) + f(w_{energy}^i) + f(w_{buff}^i) \bigg) \mathcal{Z}_i^t
 \end{align}
\begin{figure*}[!ht]
\centering
\includegraphics[scale=0.42]{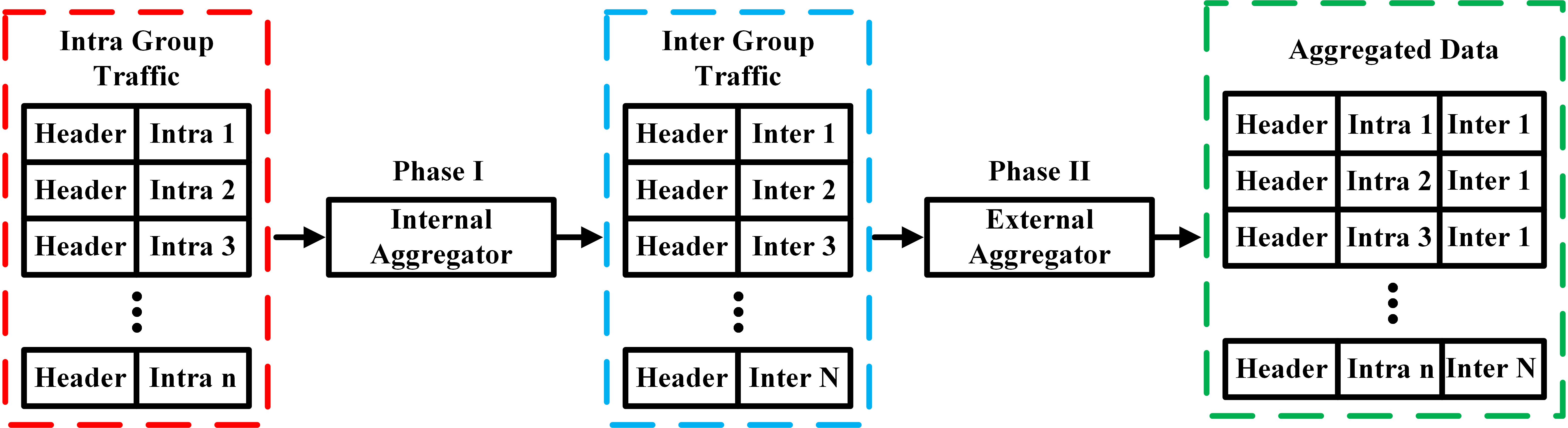}
\caption{Big data aggregation process in WBANs}
\label{fig:Agg}
\end{figure*}

\section{Cost-Effective Big Data Aggregation}
Figure \ref{fig:Agg} shows the efficient and cost-effective big data aggregation process for cloud-assisted WBANs. The process is categorized into two phases --- \textit{Phase I}: describes the data aggregation cost at LMUs for traffic flow between sensor nodes to LMUs and \textit{Phase II}: describes the data aggregation cost at BSs for traffic flow between LMUs to BSs.

\subsection{Data Aggregation: Phase I}
The big data aggregation component for the data flow between sensor nodes to LMUs is expressed as:
\begin{equation}
\mathcal{X}_{ij}^p = \mathcal{G}_{ij} \mathcal{V}_i^{k}
\end{equation}
here $\mathcal{I}_{i}^p = \sum_{j=1}^N \mathcal{X}_{ij}^p$, $\mathcal{V}_i^{k}$ represents the amount of big data originated from body sensor, and $\mathcal{G}_{ij}$ represents the resolution factor for big data transmission between IoT-enabled body sensors and LMUs. The binary decision metric for intra-BAN is expressed as:
\begin{equation}
\mathcal{G}_{ij} =\begin{cases}
     1, \quad \phantom{0}  \text{if}\,\, \text{$i$ and $j$ have a connection}  \\
     0, \quad \phantom{0}   \text{otherwise}    \\
    \end{cases}
\end{equation}  
The binary decision metric is defined to check if there is connection between a sensor node $i$ and LMU $j$. The cost incurred for aggregation data from sensor nodes to LMUs is expresses as:
\begin{equation}
\mathcal{C}_{f(w_{agg}^i)}^{intra}  = \mathcal{P}_{d_{agg}}^{intra} \mathcal{I}_{i}^p 
\end{equation}
where $\mathcal{P}_{d_{agg}}^{intra}$ denotes the intra-BAN data aggregation cost and $ \mathcal{I}_{i}^p $ denotes the intra-BAN data aggregation component.

\subsection{Data Aggregation: Phase II}
After data aggregation of physiological data at LMUs, LMUs transmit the intra-BAN medical data to BSs. Thereafter, the inter-BAN huge medical data is aggregated at BSs, which are transmitted from LMUs. Therefore, the inter-BAN big data aggregation component at BSs for the data flow between LMUs and BSs is denoted as:

\begin{equation}
\mathcal{Z}_{ij}^p = \mathcal{O}_{ij} \mathcal{F}_i^k
\end{equation} 
where $\mathcal{J}_{i}^p = \sum_{j=1}^N  \mathcal{Z}_{ij}^p$, $\mathcal{F}_i^k$ denotes the amount of data generated from LMUs, and $\mathcal{O}_{ij}$ denotes the binary decision value for the big data transfer between LMUs and BSs. The binary decision metric for inter-BAN is expressed as:
\begin{equation}
\mathcal{G}_{ij} =\begin{cases}
     1, \quad \phantom{0}  \text{if}\,\, \text{$i$ and $j$ have a connection}  \\
     0, \quad \phantom{0}   \text{otherwise}    \\
    \end{cases}
\end{equation} 
The binary decision metric is defined to check if there is connection between a LMU $i$ and BS $j$. The cost incurred for aggregation data from LMUs to BSs is expresses as:
\begin{equation}
\mathcal{C}_{f(w_{agg}^i)}^{inter}  = \mathcal{P}_{d_{agg}}^{inter} \mathcal{J}_{i}^p 
\end{equation}  
where $\mathcal{P}_{d_{agg}}^{inter}$ denotes the inter-BAN data aggregation cost and $\mathcal{J}_{i}^p$ denotes the inter-BAN data aggregation component.

\subsection{Optimization Framework Design}
Therefore the total cost inured for intra-BAN and inter-BAN big data aggregation is expressed as:
\begin{equation}
\mathcal{C}_{agg}^t = \sum_{i=1}^N  \mathcal{P}_{d_{agg}}^{intra} \mathcal{I}_{i}^p  + \sum_{i=1}^N \mathcal{P}_{d_{agg}}^{inter} \mathcal{J}_{i}^p 
\end{equation}
Now, the optimization problem for big data aggregation in intra-BAN and inter-BAN of cloud-assisted WBANs is expressed as:
\begin{align}
   &{\text{Minimize}}
   \begin{aligned}[t]
      &\sum_{t=1}^\mathcal{T} \mathbb{U}_t = \sum_{t=1}^\mathcal{T} \bigg(  \zeta_t + \sum_{i=1}^N  \mathcal{P}_{d_{agg}}^{intra} \mathcal{I}_{i}^p  + \sum_{i=1}^N \mathcal{P}_{d_{agg}}^{inter} \mathcal{J}_{i}^p \bigg)  \\
         \end{aligned} \label{Eqn:obj11} \\
   &\text{Subject to} \hspace*{1em} \sum_{t=1}^{\mathcal{T}} \mathcal{C}_{agg}^t \leq \mathcal{C}_{agg}^{th} \label{Eqn:constraint11} \\
   & \hspace*{5.3em}  f(w_{delay}^i) \geq f(w_{delay}^{th}), i \in N  \label{Eqn:constraint22} \\
   & \hspace*{5.3em}  f(w_{energy}^i) \geq f(w_{energy}^{th}), i \in N \label{Eqn:constraint33}
 \end{align}  
The intuition of this optimization problem is discussed in details. (\ref{Eqn:obj11}) illustrates the primary optimization function for data aggregation. (\ref{Eqn:constraint11}) describes that the aggregation cost, $\mathcal{C}_{agg}^t$, is to be greater than the threshold aggregation cost, $\mathcal{C}_{agg}^{th}$. The aggregation delay of the architecture, $f(w_{delay}^i)$, can not be lesser than the value of threshold aggregation delay, $f(w_{delay}^{th})$, as shown in (\ref{Eqn:constraint22}). (\ref{Eqn:constraint33}) represents that the energy of big data-enabled WBANs, $f(w_{energy}^i)$, can not be lesser than the value of threshold energy, $f(w_{energy}^{th})$. We solve the optimization problem using Lagrangian Multipliers to optimize aggregation cost and delay.

\begin{table}[!h]
    \centering
    \caption{Experimental Parameters}
    \vspace{0.1cm}
    \scalebox{0.78}{
    \begin{tabular}{l|c}
        \hline
        \multicolumn{1}{c|}{\textbf{Parameter}} & \multicolumn{1}{c}{\textbf{Value}} \\
        \hline
        \hline
        Experimental area & {3.5 Km $\times$ 3.5 Km} \\
        Experimental time & 1 hour \\
        Count of WBANs & 400 \\
        Count of IoT-enabled body sensors & 8 \\
        Remaining energy of a WBAN & 0.5 J \\
        WBAN mobility model & 1.5-2.5 m/s \\
        Transmit energy expenditure  & 16.7 nJ \\
        Receive energy expenditure & 36.1 nJ \\
        Amplifier energy expenditure & 1.97 nJ \\
        Body sensor communication range  & 0.5-1.5 m \\
        Data incoming rate & 6 packets/sec \\
        Data size & 1 Mb \\
        \hline
    \end{tabular}
    }
    \label{tab:sim}
\end{table}

\begin{table}[!h]
    \centering
    \caption{Placement position of IoT-enabled body sensors \cite{Cui2013}}
    \vspace{0.1cm}
    \scalebox{0.78}{
    \begin{tabular}{l|c|c}
    \hline
\multicolumn{1}{c|}{\textbf{Sensor node}} & \multicolumn{1}{|c|}{\textbf{Data rate}} & \multicolumn{1}{c}{\textbf{Position}} \\
        \hline
        \hline
        1) ECG node     & 71 kbps   & ($15,78$) \\
        2) Motion node  & 35 kbps   & ($20,52$) \\
        3) EEG node     & 43.2 kbps & ($57,90$) \\
        4) Glucose node & 1.6 kbps  & ($44,108$)  \\
        5) EMG node     & 100 kbps  & ($17,99$)   \\
        6) EMG node     & 100 kbps  & ($22,120$)  \\
        7) Motion node  & 35 kbps   & ($34,0$)    \\
        8) Motion node  & 35 kbps   & ($50,0$)   \\
        \hline
    \end{tabular}
    }
    \label{tab:tab2}
\end{table}
\section{Performance Analysis}
To analyze the performance of proposed approach -- DEBA, we enlisted the experimental settings in Table \ref{tab:sim}. 

\subsection{Experimental Settings}
We take into consideration an one-hop topological framework for the data transfer between body sensors to LMUs and LMUs to BSs. We configure an experimental setting, where we consider $400$ WBANs in a span of area 3.5 Km $\times$ 3.5 Km and the WBAN is composed of $8$ big data-enabled body sensors. The data transmission rate of IoT-enabled body sensors changes within rage of $8$ Kbps to $8$ Mbps as described in Table \ref{tab:tab2}. The WBANs communicate with $15$ BSs in a particular area through LMUs. We consider the Group-based mobility of WBANs to configure the mobility of WBANs. We consider the energy consumption model as discussed in \cite{samanta2015link}. On the other hand, we consider the Raleigh fading for the data transmission between IoT-enabled body sensors and LMUs with path loss exponent changes within a range of $2 - 3.2$. On the other hand, we also take into consideration the Log-normal fading technique for the big data transfer between LMUs and BSs with path loss exponent changes within a rage of $3.5 - 4$ \cite{Sun2015, Shen2019}. We configure the packet size of $1$ Mb for the communications in Phase I and II. The maximum transmit energy expenditure of a transmitter is set to 55 nW (-42.6 dBm) and 12 nW (-49.2 dBm) for both mobile and static WBANs. The experimental setup follows all other settings of the IEEE 802.15.6 standard \cite{IEEE_Std} to determine the experimental results of DEBA in real environment. The criticality factor of WBANs varies within the span of $0-1$. The WBANs with criticality factor greater than $0.5$ is considered to be critical, otherwise the WBAN is considered to be normal \cite{Shimly2019}. 

\begin{figure}[ht!]
\centering
\subfigure[Aggregation Delay]{%
\includegraphics[scale = .24]{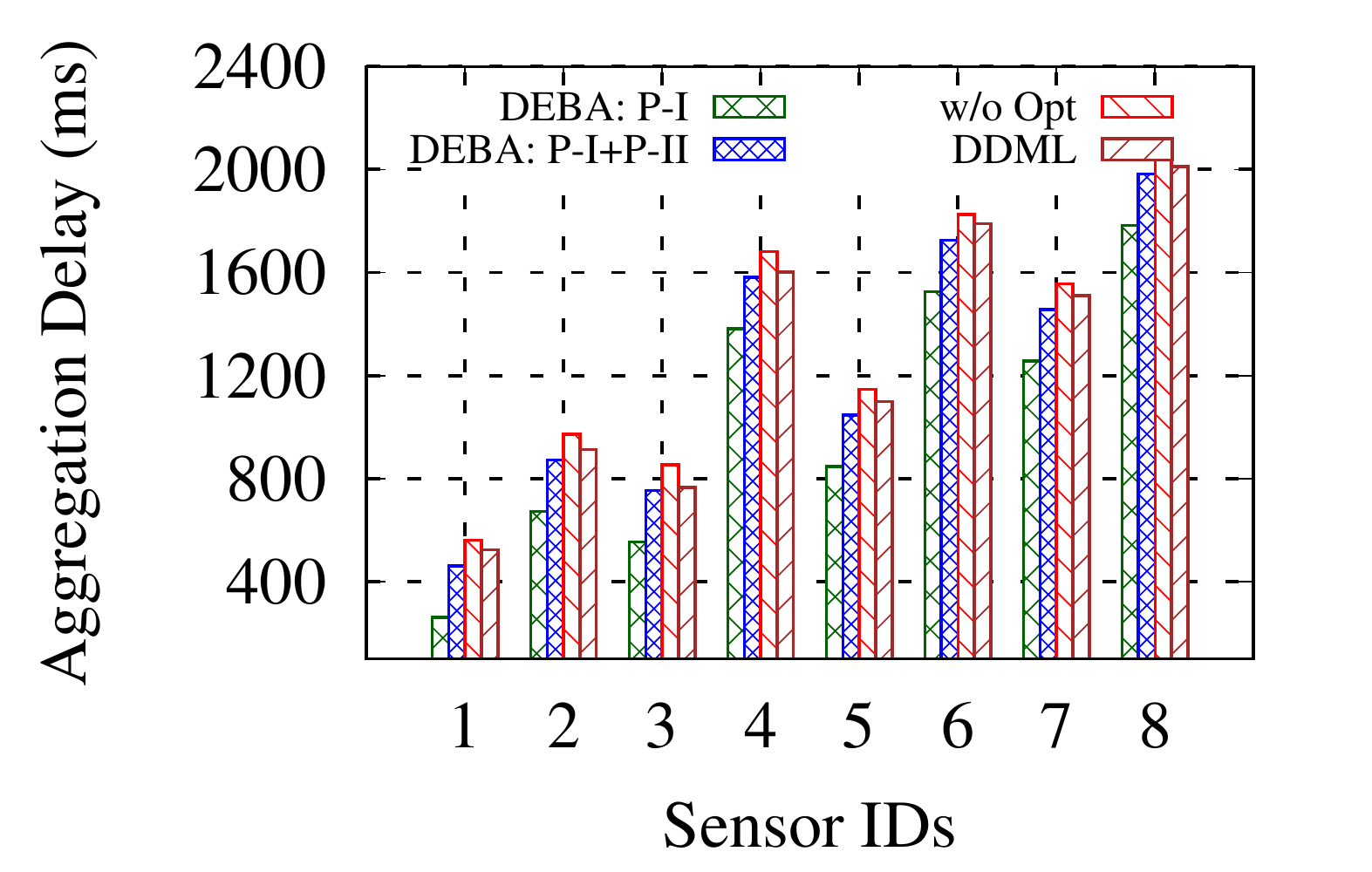}
\label{delay}}
\quad \hspace{-8mm}
\subfigure[Aggregation Cost]{%
\includegraphics[scale = .235]{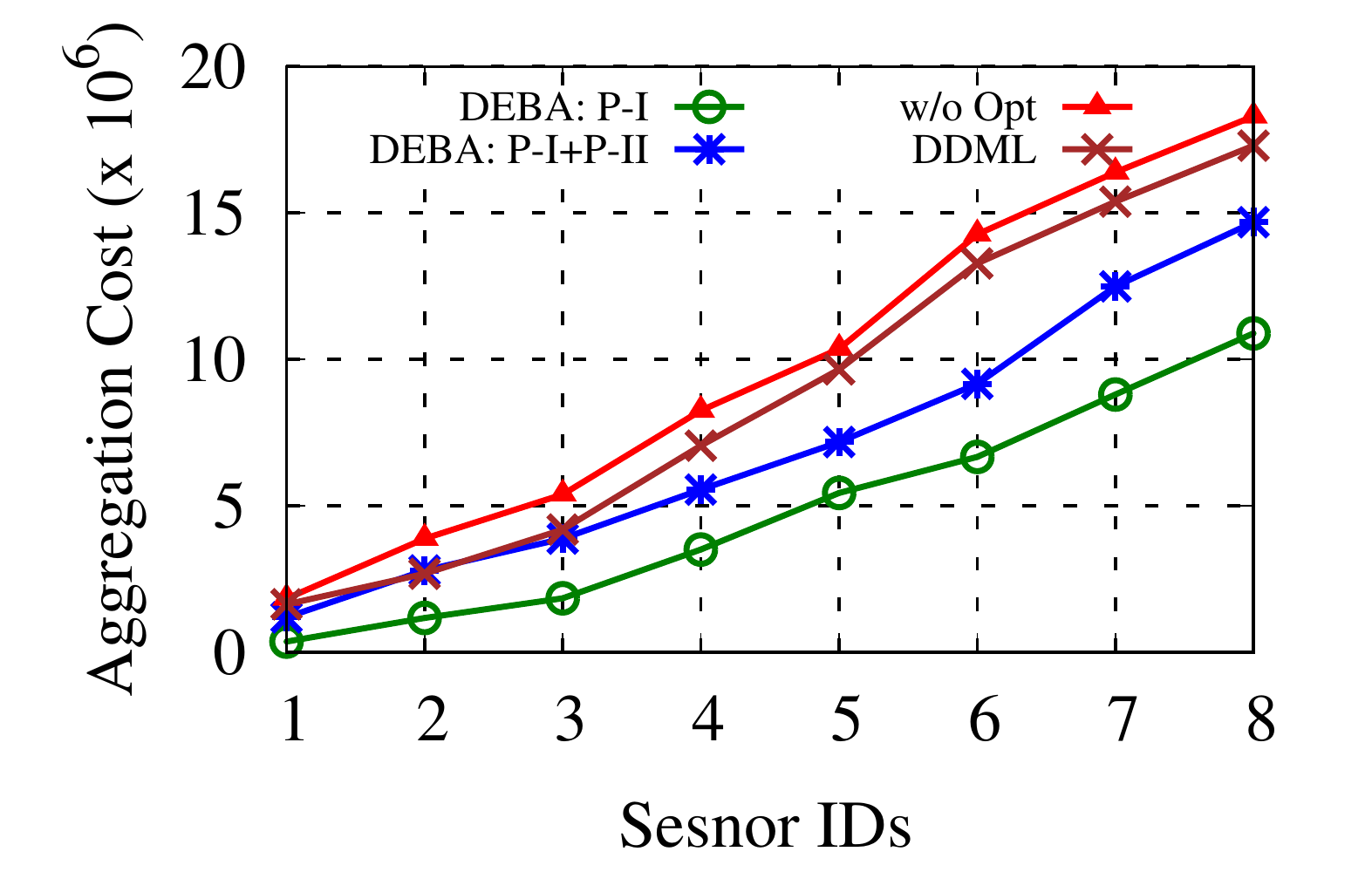}
\label{Network_cost}}
\caption{Analysis of data aggregation delay and aggregation cost with P-I, P-I+P-II, DDML and without optimization.}
\label{a1}
\end{figure}
\subsection{Discussion on Experimental Results}
Figure \ref{delay} illustrates the big data aggregation delay of big data-enabled WBANs. We compared DEBA with DDML\footnote{DDML is data dissemination technique for WBANs. It used machine learning techniques to optimize the data dissemination technique for WBANs.} \cite{punj2022data}. The figure represents that the DEBA: Phase I (P-I) and DEBA: Phase I+Phase II (P-I+P-II), the data aggregation process incurs lesser delay than the approach without any optimization (i.e., w/o Opt) and DDML. As the aggregation delay is optimized for cloud-assisted WBANs in the data aggregation function, as shown in Equation \ref{main}. Figure \ref{Network_cost} shows the data aggregation cost for cloud-assisted WBANs.

\begin{figure}[ht!]
\centering
\subfigure[Traffic Served]{%
\includegraphics[scale = .245]{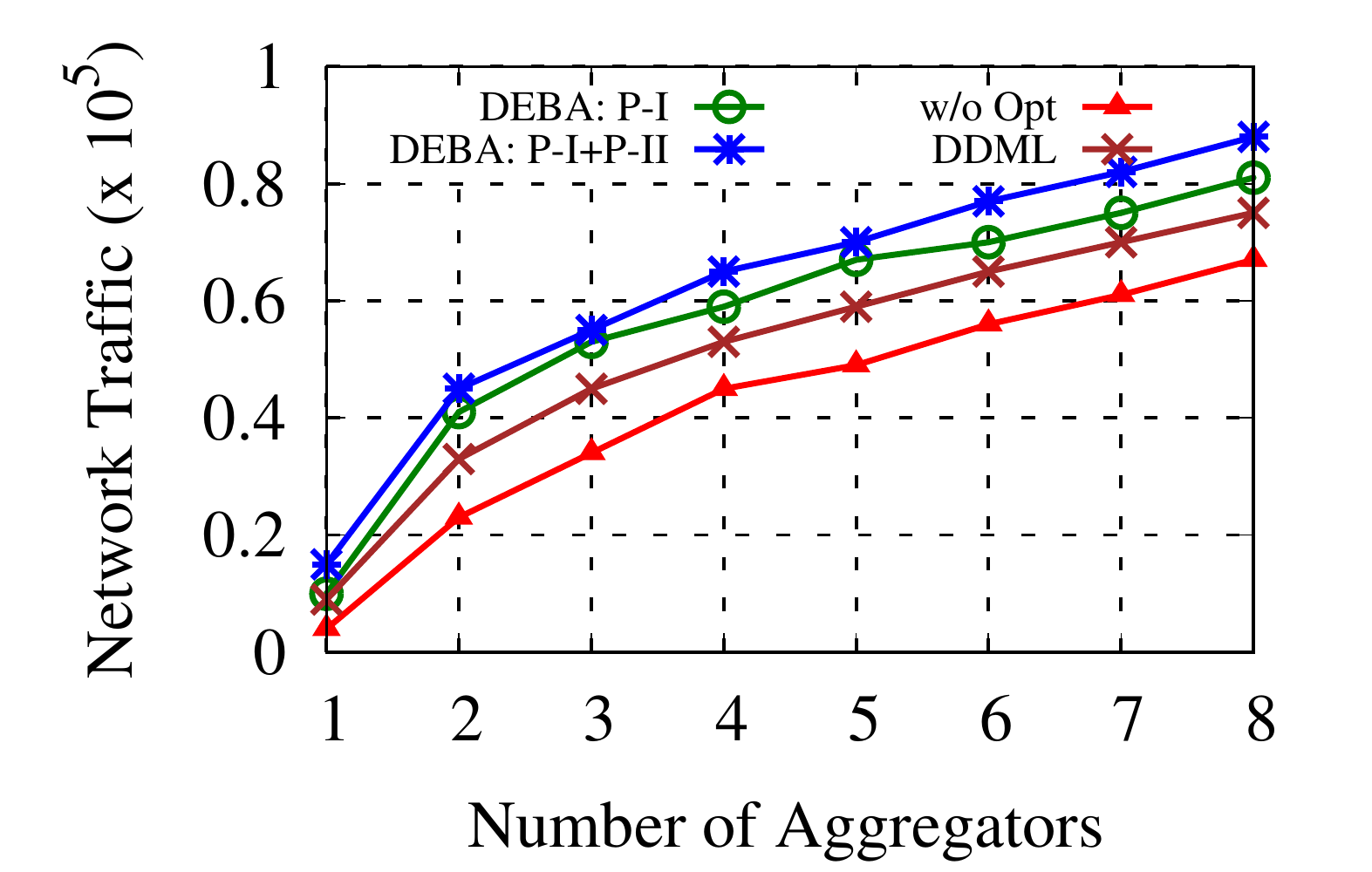}
\label{traffic}}
\quad \hspace{-8mm}
\subfigure[Energy Consumption]{%
\includegraphics[scale = .24]{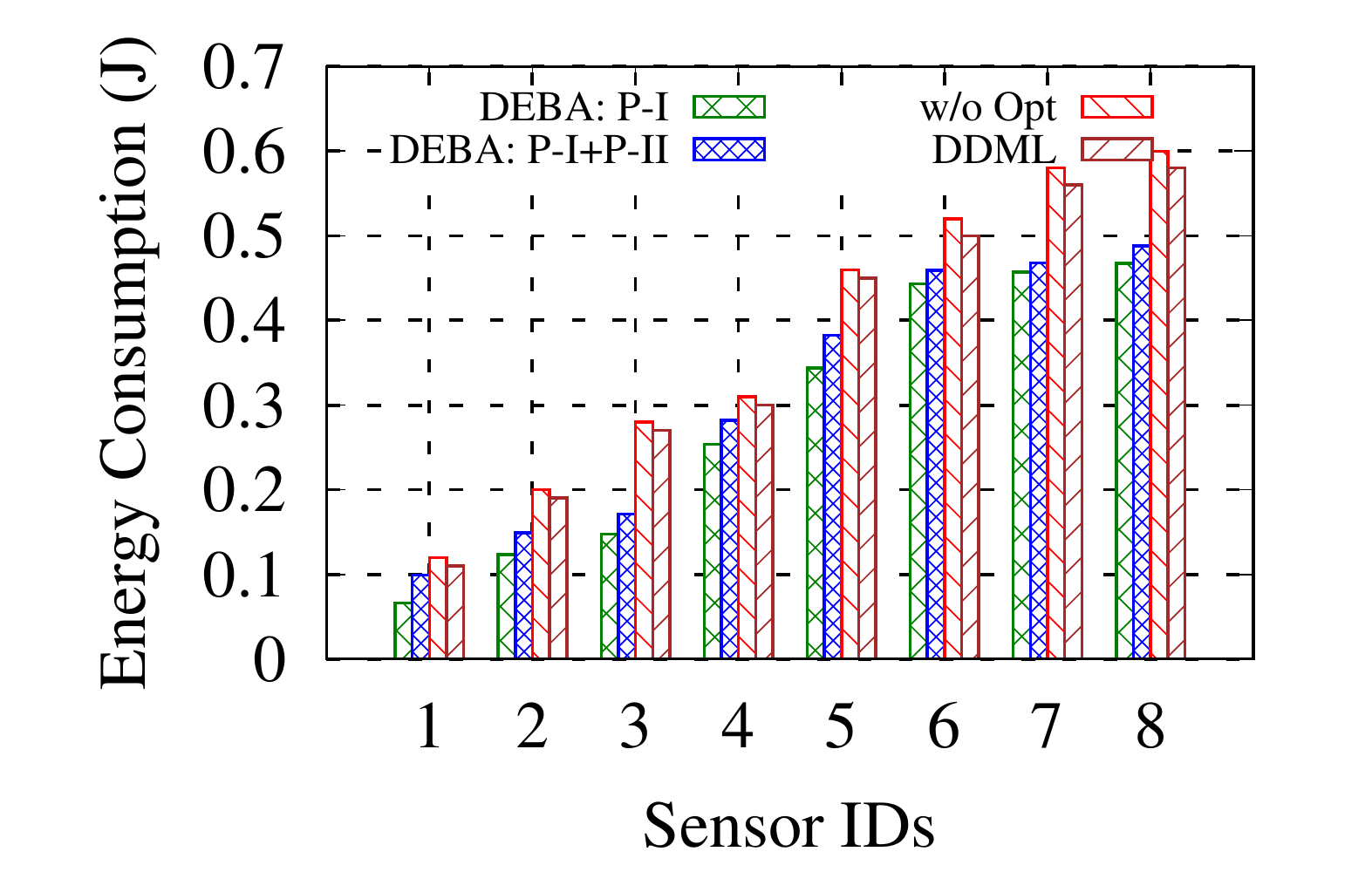}
\label{energy}}
\caption{Analysis of traffic served and energy consumption with P-I, P-I+P-II, DDML and without optimization.}
\label{a2}
\end{figure}
We see that the DEBA optimizes the data aggregation cost and performs better than the previous approach DDML. Figure \ref{traffic} illustrates the network traffic for the aggregation process in cloud-assisted WBANs. The figure illustrates that the traffic served for big data-enabled WBANs is less without any optimization technique than DEBA: P-I and DEBA: P-I+P-II. Using DEBA, we gain 5-7$\%$ improvement in the performance in terms of traffic served. Consecutively, Figure \ref{energy} shows the energy consumption of sensor nodes for DEBA. The figure represents that DEBA provides 7-8$\%$ improvement in terms of energy consumption. Therefore, DEBA performed better than the existing approach DDML and w/o Opt.

\section{Conclusion}
In a pandemic situation, the data creation rate from big data-enabled body sensors expands significantly, which minimizes the energy efficiency of sensor nodes and maximizes the big data aggregation cost for big data-enabled WBANs. Therefore, we proposed a quality-driven and energy-efficient big data aggregation approach for WBANs. We theoretically model the proposed approach --- {\tt DEBA} and it provides a significant improvement in terms of data aggregation delay, aggregation cost, and energy efficiency. 

\bibliographystyle{IEEEtran}
\bibliography{1st}

\end{document}